# Efficient Dynamic Nuclear Polarization at High Magnetic Fields

Gavin W Morley and Johan van Tol
Center for Interdisciplinary Magnetic Resonance, National High Magnetic Field Laboratory at Florida State University, Tallahassee, FL 32310, USA

Arzhang Ardavan
Clarendon Laboratory, Department of Physics, University of Oxford, OX1 3PU, U.K.

Kyriakos Porfyrakis, Jinying Zhang, and G. Andrew D. Briggs
Materials Department, University of Oxford, OX1 3PH, U.K.

By applying a new technique for dynamic nuclear polarization involving simultaneous excitation of electronic and nuclear transitions, we have enhanced the nuclear polarization of the nitrogen nuclei in $^{15}$N@C$_{60}$ by a factor of $10^3$ at a fixed temperature of 3 K and a magnetic field of 8.6 T, more than twice the maximum enhancement reported to date. This methodology will allow the initialization of the nuclear qubit in schemes exploiting N@C$_{60}$ molecules as components of a quantum information processing device.

The incarceration of atomic nitrogen in a C$_{60}$ cage leads to a species (known as N@C$_{60}$) with remarkable properties: The nitrogen occupies a high-symmetry site at the center of the cage and retains its atomic configuration, and the cage offers protection of the nitrogen electron paramagnetic moment from interactions with the environment[1]. It exhibits the longest electron spin lifetimes observed in any molecular species with a phase coherence time $T_2 = 240$ $\mu s$ [2] in solution at 170 K, and spin-lattice relaxation time, $T_1$, of 4.5 minutes at 4 K[3]. $T_2$ is of the order of $10^4$ times longer than the time taken to manipulate the electron spin state by pulsed electron spin resonance (ESR), and this has led to speculation that it may be a useful component in an electron-spin-based quantum computer[4,5,6,7,8].

The nuclear spin of the incarcerated N atom exhibits even longer lifetimes, and it too can in principle be used to store quantum information. However, before the nuclei can be exploited as qubits, a method of polarizing them would be required, and this method should work at temperatures that are high on the scale of the nuclear Zeeman energy. (While the preparation of pseudo-pure initial states allows high-temperature NMR quantum computing[9], the readout signal becomes exponentially small as the number of qubits goes up[10]). Methods of achieving nuclear polarization are also useful in other contexts, including NMR[11] and high-energy physics[12].

Four microwave-induced dynamic nuclear polarization (DNP) effects have been demonstrated in the past: The Overhauser effect (OE)[13], the solid effect (SE), thermal mixing (TM) and the cross effect (CE). These are described in reference [14] and the articles cited therein. Defining the polarization enhancement as $\varepsilon = \dfrac{P_{\text{after DNP}}}{P_{\text{before DNP}}}$ (where the polarization, *P*, is the net number of aligned nuclear spins divided by the total number), these techniques lead to enhancements in the NMR signal intensity of up to $\varepsilon = 400$ [11]. In this Letter we describe a new approach in which the nuclei are polarized using electron nuclear double resonance (ENDOR), exploiting the large difference between the electronic and nuclear spin-lattice relaxation times. We refer to this technique as 'polarization of nuclear spins enhanced by ENDOR' (PONSEE). PONSEE has several significant advantages over SE, TM and CE DNP. First, the polarization enhancement for PONSEE increases with magnetic field, while SE, TM and CE DNP all produce less polarization as the field increases. This makes PONSEE well-suited for use in contemporary NMR spectrometers which benefit from the improved spectral resolution available at 7-21 T. Second, the time required for PONSEE is a few times the electronic $T_1$, which is shorter than the OE, SE, TM and CE because no forbidden transitions or three-spin transitions are used.

Room-temperature NMR signal-to-noise enhancements of >10,000 have been achieved by cooling a sample, performing 'traditional DNP', and then (in a time that is short compared to the nuclear $T_1$) re-warming [15]. The 'traditional DNP' created a polarization of 26% with one hour of the TM effect at 1.1 K, but using PONSEE instead of TM would have produced larger polarization in a shorter time.

A molecule of $^{15}$N@C$_{60}$ in a strong magnetic field has eight energy levels as shown in Fig. 1. The electronic and nuclear spins, $S_z$ and $I_z$, are both good quantum numbers. The $^{15}$N isotope has spin **I** = ½ which is a natural qubit unlike the more abundant $^{14}$N nucleus which has **I** = 1. Complete two-qubit quantum computation is feasible with an ensemble of $^{15}$N@C$_{60}$[3].

At a temperature of $T = 3$ K and a magnetic field of $B = 8.6$ T, the electronic Boltzmann factor

$$\alpha = \frac{g\mu_B B}{k_B T} = 0.02. \tag{1}$$

This implies that about 98% of the electrons are in the $S_z = -3/2$ state and only 0.001% are in the $S_z = 3/2$ state.

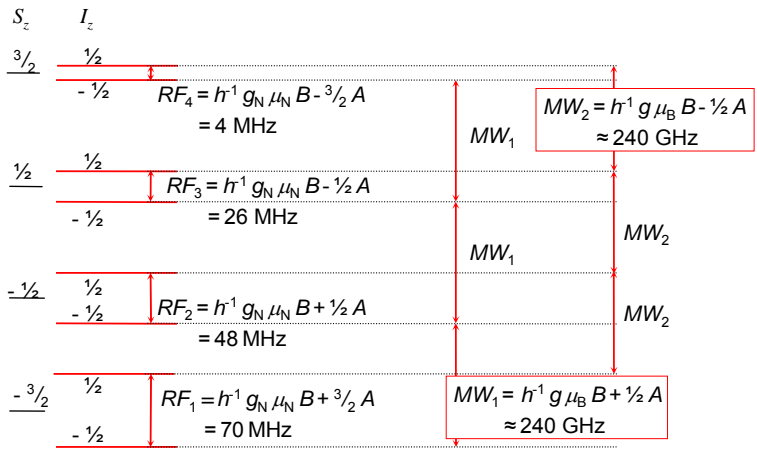

**Fig. 1. (Color Online) Energy levels and allowed transitions of $^{15}$N@C$_{60}$ in a strong magnetic field, $B$, parallel to the $z$-axis. The system is well described by the effective spin Hamiltonian $H = B(g\mu_B S_z - g_N \mu_N I_z) + hAS_z I_z$. The three terms describe the electronic Zeeman, nuclear Zeeman and isotropic hyperfine interactions. The hyperfine constant is $A = 21.9$ MHz. The four nuclear transitions are at radio frequencies (RF) so are labeled $RF_{1-4}$. The two triply-degenerate electronic transitions are at microwave frequencies so are labeled $MW_{1,2}$.**

Our PONSEE technique consists of the three steps shown in Fig. 2. The first (second) step selectively excites electronic (nuclear) spins whose nuclear (electronic) spins are in a particular state. The third step requires that $T_1^{\text{nuclear}} \gg T_1^{\text{electronic}}$, which is generally true for all spin systems.

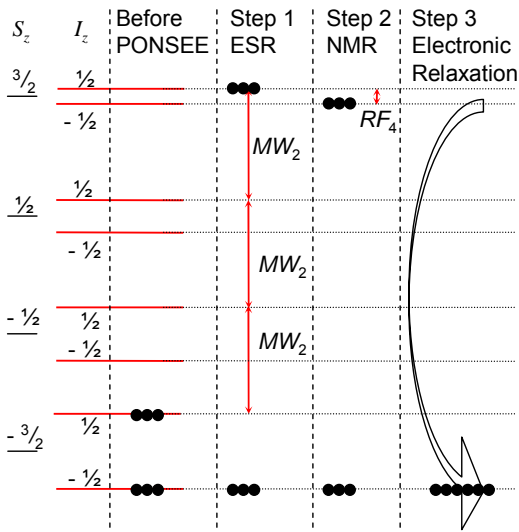

**Figure 2. (Color online) Schematic of the pulsed version of the PONSEE sequence for $^{15}$N@C$_{60}$. $RF_4$ radiation is shown in step 2, but $RF_3$ or $RF_2$ could be used instead.**

PONSEE works efficiently with either continuous-wave (CW) or pulsed radiation. With CW radiation, the three steps take place simultaneously. If pulsed radiation is used, $\pi$ pulses should be chosen and after four applications of the three-step cycle, the nuclear polarization reaches the CW result for $\alpha \ll 1$[3].

PONSEE is distinct from electron-nuclear cross polarization (eNCP)[14] which also uses ESR and NMR excitations, in that eNCP relies on spin-locking which operates in pulsed mode only.

Our experiments used CW radiation for the first two steps. To maximize the nuclear polarization, the radiation should be powerful enough to saturate the transitions, equalizing the populations of the connected energy levels. RF radiation with frequency $RF_4$ accesses the maximum electronic polarization available for a given combination of temperature and magnetic field. The polarization available with $RF_3$ and $RF_2$ radiation is equal to the lesser electronic polarization of the $S_z = -3/2$ state with respect to the $S_z = \frac{1}{2}$ and $S_z = -\frac{1}{2}$ states respectively.

$^{15}$N@C$_{60}$ was prepared by ion implantation[16] with isotopically pure $^{15}$N and purified with high performance liquid chromatography (HPLC)[17]. HPLC, combined with ESR and UV characterization indicated that about 80% of the cages contained nitrogen atoms. The purified sample was dissolved in deuterated toluene and put into a quartz ESR tube of outer diameter 4 mm. The solvent was deoxygenated with three freeze-pump-thaw cycles and sealed under a dynamic vacuum.

Measurements were made with a home-built 240 GHz CW ESR spectrometer[18]. Fig. 3 shows an ESR spectrum recorded at 3 K. The absorption lineshape is characteristic of a saturated electronic spin-system with a long spin-lattice relaxation time, $T_1$[19], and shows no nuclear polarization.

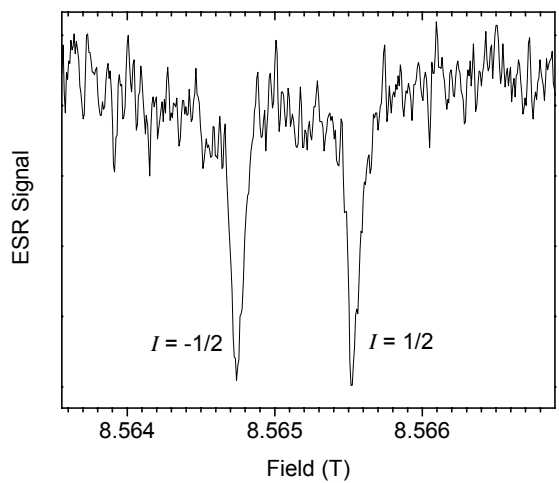

**Figure 3. 240 GHz CW ESR spectrum of $^{15}$N@C$_{60}$ in thermal equilibrium at 3 K.**

A small amount of nuclear polarization can be produced by applying microwave radiation only. Fig. 4 shows the ESR spectrum after radiation of frequency $MW_2$ was applied for 2.5 hours at 4.2 K. The relative area of the two peaks shows that the nuclear polarization is 21 ±2%. This form of DNP is due to the Overhauser effect[13].

Repeating this experiment with $MW_1$ radiation did not produce anti-polarization. This shows that the polarization in Fig. 4 does not result from a difference in the $T_1$ times of the four nuclear transitions. If the nuclear $T_1$ time of the $RF_1$ transition was much longer than that of $RF_2$, $RF_3$ or $RF_4$, then it would be possible to polarize or anti-polarize the nuclear spins by exciting the electronic spins associated with anti-aligned or aligned nuclear spins respectively.

Instead, the polarization achieved with only $MW_2$ radiation is due to flip-flop transitions in which the nuclear and electronic spins simultaneously flip in opposite directions, conserving total angular momentum. When an electron relaxes back to a lower energy level it loses a quantum of angular momentum, so the corresponding part of a flip-flop transition must add a quantum of angular momentum to the nuclear spin. This effect always acts to align nuclear spins with the magnetic field, and occurs more often than the competing transitions in which the electron and nucleus must both lose a quantum of angular momentum. The perturbation causing flip-flops arises from the non-secular part of the isotropic hyperfine coupling, as can be seen when this interaction is expressed in terms of the raising ($S^+$ and $I^+$) and lowering ($S^-$ and $I^-$) operators:

$$A \mathbf{S} \bullet \mathbf{I} = \frac{A}{2}(S^+ I^- + S^- I^+) + A S_z I_z. \tag{2}$$

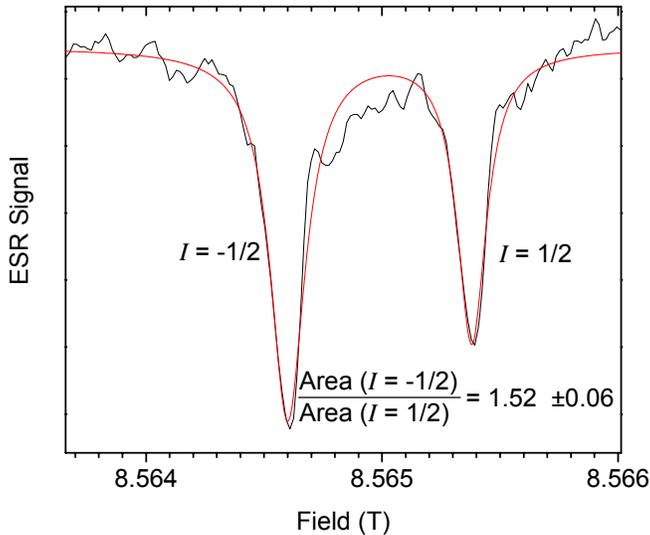

**Figure 4. (Color online) CW ESR spectrum of $^{15}$N@C$_{60}$ at 4.2 K after 2.5 hours $MW_2$ radiation. The nuclear polarization is due to flip-flop transitions. The smooth curve is a fit to two Lorentians. The ratio of the areas of these peaks shows the nuclear polarization to be 21%.**

We found that the DNP effect was stronger with PONSEE than with the Overhauser effect. Furthermore, it was equally possible to align the nuclear spins with or against the applied field using PONSEE. Applying $MW_2$ and $RF_2$ radiation for 20 minutes produced the spectrum shown in Fig. 5 in which the nuclear polarization is 62 ±2%. In theory, exciting the $RF_3$ and $RF_4$ transitions should polarize the nuclear spins more strongly, but instead we found that the effect was weaker. We attribute this to the lower

RF power available at these frequencies, and less effective (frequency-dependent) coupling to the sample in our apparatus.

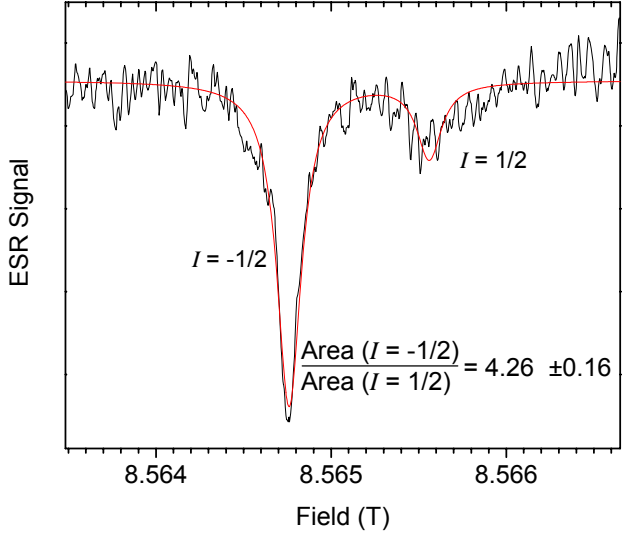

**Figure 5. (Color online) CW ESR spectrum of $^{15}$N@C$_{60}$ at 3 K after 20 minutes PONSEE. The smooth curve is the bi-Lorentzian best-fit with the linewidth of the two peaks constrained to be the same. The nuclear polarization is 62%, an enhancement of $\varepsilon = 1100$.**

The thermal equilibrium polarization of isolated $^{15}$N nuclear spins at 3 K in a field of 8.6 T is 0.03%[20]. Confinement inside a C$_{60}$ cage increases the strength of the hyperfine interaction, so the corresponding equilibrium polarization in $^{15}$N@C$_{60}$ is 0.06%. Our PONSEE experiment produced 62% nuclear polarization, an enhancement of $\varepsilon = 1100 \pm 50$.

We found that the $RF_2$ transition was excited most effectively by sweeping the RF frequency from 47.916 to 47.924 MHz about once a minute. This frequency range provides a measurement of the magnitude of the nuclear $g$-factor via the expression given in Fig. 1:

$$|g_N| = \frac{h(RF_2 - A/2)}{\mu_N B} = 0.566(2), \tag{3}$$

using the value of the hyperfine constant measured from Fig. 4, $A = 21.9 \pm 0.15$ MHz[a]. The accepted value of $g_N$ for $^{15}$N is -0.566[13], in agreement with our measurements.

The frequencies of the other nuclear transitions were calculated with this nuclear $g$-factor. The measured transition frequencies agree with these calculations as shown in Table 1.

---

[a] Our hyperfine constant agrees with that measured at the lower magnetic field of 0.34 T [1].

| Nuclear Transition | Frequency Calculated from $RF_2$ Transition (MHz) | Measured Frequency (MHz) | Measuring Technique |
|---|---|---|---|
| $RF_1$ | $69.80 \pm 0.13$ | $69.875 \rightarrow 69.900$ | Depolarization * |
| $RF_2$ | N/A | $47.916 \rightarrow 47.924$ | DNP |
| $RF_3$ | $26.03 \pm 0.04$ | $25.9 \rightarrow 26.1$ | DNP (weak) |
| $RF_4$ | $4.15 \pm 0.13$ | $3.8 \rightarrow 4.8$ | DNP (weaker) |

**Table 1. Comparison of measured and calculated frequencies of nuclear transitions in $^{15}$N@C$_{60}$. The transitions are labeled in Fig. 1. The uncertainties in the calculated frequencies are due to the uncertainty in the measurement of the hyperfine constant. The range in the measured frequencies corresponds to the RF sweep range used. Nuclear transitions for which the applied RF is more strongly coupled to the sample have a stronger effect over a smaller frequency range. *The $RF_1$ transition frequency was measured by performing a series of `depolarizing' experiments after DNP. These experiments consisted of applying $RF_1$ radiation only for about a minute, partially equalizing the sizes of the two resonances.**

A simple model was used to predict the amount of polarization observed, assuming complete saturation of transitions $MW_2$ and $RF_2$. The nuclear polarization is

$$P_{\text{after PONSEE}} = \frac{1/\alpha - 3 + \alpha + \alpha^2}{1/\alpha + 5 + \alpha + \alpha^2} = 84\%, \qquad (4)$$

for $\alpha = 0.02$. The measured polarization of 62% is smaller than this for two reasons: The RF radiation raised the sample temperature, and the $RF_2$ transition was not saturated. The RF power was set to 40 W to strike a balance between these competing issues. If the $RF_4$ transition had been used, our model predicts a nuclear spin polarization of 99.99%, corresponding to $\varepsilon = 1700$.

Once a nuclear polarization has been achieved, it is possible to investigate the nuclear $T_1$. After producing a 62% nuclear polarization, the temperature was raised to 4.2 K and the sample was allowed to relax for 11.5 hours. The spectrum recorded after this time revealed that the nuclear polarization had only fallen to 40%. More measurements would be needed to properly describe the nuclear spin relaxation of $^{15}$N@C$_{60}$, but these data show that the nuclear $T_1$ is on the order of twelve hours at 4.2 K. This long time means that $^{15}$N nuclear spin flips are not a significant mechanism for the electronic $T_2$ relaxation of $^{15}$N@C$_{60}$ at this temperature.

In summary, we have demonstrated a new DNP technique, PONSEE, achieving a polarization enhancement exceeding $\varepsilon = 10^3$ in a sample of $^{15}$N@C$_{60}$ at the high magnetic field of 8.6 T. PONSEE will be a valuable tool in initializing the nuclear qubits in quantum information processing devices employing N@C$_{60}$ molecules; the efficiency of this technique is independent of the number of qubits involved in the computation. Our measurement of a very long nuclear $T_1$ time suggests that these nuclear states could be useful as memory elements. The long electronic $T_1$ offers the promise of a long nuclear qubit coherence time, insofar as electron spin flips provide a decoherence mechanism. Furthermore, PONSEE could find applications beyond quantum information processing; NMR experiments at high fields could benefit from signal-to-noise enhancements to selected nuclei using PONSEE.